\documentclass[showpacs,preprintnumbers,amsmath,amssymb]{revtex4}

\usepackage{amssymb}

\begin{document}

\setlength{\textheight}{23cm}
\setlength{\textwidth}{17cm}
\setlength{\oddsidemargin}{-.5cm}
\setlength{\topmargin}{-1cm}

%\numberwithin{equation}{\arabic{section}}
 \makeatletter
\@addtoreset{equation}{section}
\renewcommand{\theequation}{% \theequation '̍āX'è‹`
  \arabic{section}.\arabic{equation}}
\makeatother

%%%%%%%%%%%%%%%%%%%%%%%%%%%%%%%%%

\title {A thermodynamical model  for  
non-extremal black $p$-brane }
\date{\today} 
\author{Ken-ichi Ohshima } \email []{ohshimak@hep1.c.u-tokyo.ac.jp}
\affiliation{Institute of Physics, University of Tokyo \\
 Komaba, Meguro-ku 153 Tokyo}

\thispagestyle{empty}\setcounter{page}{0}

\begin{abstract}
We show that the correct entropy, temperature (and absorption probability)
of non-extremal black $p$-brane can be reproduced by a certain 
thermodynamical model when maximizing its entropy. We show that
the form of the model is related to the geometrical similarity of 
non-extremal and near extremal black $p$-brane at near horizon region,
and argue about the appropriateness of the model.
\end{abstract}

%%%%%%%%%%%%%%%%%%%%%%%%%%%%%%%%%%%%%%%%%%%%%%%%%%%%%%%%%%%%% 
\maketitle 
%%%%%%%%%%%%%%%%%%%%%%%%%%%%%%%%%%%%%%%%%%%%%%%%%%%%%%%%%%%%% 

%%%%%%%%%%%%%%%%%%%%%%%%%%%%%%%%%
%
%
%   Introduction 
%
%
%%%%%%%%%%%%%%%%%%%%%%%%%%%%%%%%%

\section{Introduction}
\ \ \ \ \ The microscopic origin of black hole entropy has been considered to be 
an important subject for years,  since the mechanism of black hole thermodynamics 
is regarded to  be explained by quantum gravity theory.  String theory, 
the leading candidate of quantum gravity theory, actually explained a
black hole entropy by a microscopic model \cite{SV}.  A certain D-brane model
exactly reproduces the black hole entropy including its coefficient.
It is based on   supersymmetric nature of the black hole,
which ensures vanishing quantum corrections so that the string - D-brane model 
defined at weak coupling region reproduces entropy in
strong coupling region.   Due to similar supersymmetric nature,
AdS/CFT correspondence \cite{ads} has been providing various important results.

 On the other hand, when supersymmetry is broken, the relation between D-brane
model in weak coupling region and supergravity (of strong coupling 
region) is unclear, since the quantum correction is unknown.
It is difficult to  relate  string theoretical models to
black hole thermodynamics in general.   However, it is important to look for
thermodynamical models which can describe black hole thermodynamics, 
since one might find a clue to understand non-perturbative 
property of quantum gravity theory. 

For example, \cite{B3} showed that
D3-brane and open string gas model provides the microscopic description of 
near extremal black 3-brane thermodynamics, up to numerical factor.
The reason of this agreement without supersymmetry is explained in  \cite{Mal} \cite{das1} \cite{das},
and the reason of the discrepancy of the coefficient is studied in \cite{sia}\cite{atl}. 
In addition, \cite{Bp} showed that  the entropy of  near extremal 
black $p$-brane is described as $S \propto q^a T^b$ where $T$ is the temperature and $q$ is the charge.

  Surprisingly, \cite{DGK}  showed that D3-brane - anti D3-brane model also 
provides the microscopic description of non-extremal (and in particular Schwarzschild)  
thermodynamics, including its absorption probability \cite{abs8}, up to numerical factor. 
\cite{SP} \cite{BL} extended it to general neutral $p$-brane and showed 
that the similar model reproduces the correct entropy up to numerical factor. 
It is also applied to non-extremal black $p$-branes, multi-charged branes, rotating branes 
\cite{Kal1} \cite{l} \cite{KalS} \cite{ght} \cite{Halyo}.  Recently, \cite{braneant}
showed a relation  between the thermodynamics of chargeless black 3-brane and 
D3-brane - anti D3-brane system, in the supergravity framework. 

 All those thermodynamical models for black 3-brane are $T^4$ model, i.e., 
the thermodynamics of black 3-brane is 
described like $\mathbb{R}^{3+1}$ massless field in finite temperature.
 This might indicate a correspondence between gravity and other massless theory. 
In addition, all those models are  ``extended near extremal model'', i.e., 
when the thermal energy of the near extremal $p$-brane is described by $T^\alpha$, 
that of  the non-extremal $p$-brane (far from extremality) is also 
described by  $T^\alpha$.

 In this paper, we propose an expression of a ``phenomenological'' thermodynamical model which 
provides the correct entropy and temperature and absorption probability 
including coefficients.  This expression does not yield additional discrepancy in
the coefficient, and exactly reproduces the supergravity results.
Next we show a geometrical similarity  between the near extremal black $p$-brane and
non-extremal black $p$-brane.  This geometrical similarity
results in the ``extended near extremal model'' 
which we mentioned above. We also show that our thermodynamical model is 
in fact appropriate for blackbody radiation of non-extremal black 3-brane, by considering
the action of  dilaton field in the low energy region.  
And we show that the correct absorption probability and greybody factor
for non-extremal black 3-brane are also obtained. We show that this exact agreement is due to the
equality of the scalar field equation in black 3-brane background and 
parameter changed near extremal 3-brane background.   

  The organization of this paper is as follows.  
In section \ref{blackpreview}, we briefly review   black $p$-brane and 
the thermodynamic quantities of black $p$-brane. and the thermodynamical models
proposed in the past.
In section \ref{honron}, we explain our thermodynamical model ansatz. After introducing
the ansatz, we show a geometrical similarity between  non-extremal black $p$-brane and
near extremal black $p$-brane.  We also show that the agreement of the temperature,
the entropy and the absorption probability is related to the geometrical similarity.  
We also argue about the appropriateness of our ansatz in terms of the geometrical similarity
and the action of massless scalar field.
Section \ref{lastsection} is the conclusion and discussion.

%%%%%%%%%%%%%%%%%%%%%%%%%%%%%%%%%
%
%
%   Black p-brane 
%
%
%%%%%%%%%%%%%%%%%%%%%%%%%%%%%%%%%

\section{Black $p$-brane thermodynamics and thermodynamical models \label{blackpreview}}

\subsection{Black $p$-brane and its thermodynamical quantities \label{blackthermo}}

\ \ \ \ The black $p$-brane solution in 10 dimensional supergravity 
of 1/2 BPS in the extremal limit 
is described as extrema of the following supergravity action 
\cite{Bp}\cite{bp1}\cite{bp2}\cite{bp3}\cite{bp4}\cite{bp5}\cite{bbd}\cite{spb} 
\begin{equation}
 S \  = \ - \dfrac{1}{2 \kappa^2 } \int d^{10}x \sqrt{-g}
\left( R - \dfrac{1}{2} (\partial \phi )^2 - \dfrac{1}{ 2 (8-p)! } e^{a \phi } {F_{8-p}}^2 \right) \ \ ,
\label{sugraaction}
\end{equation}
the metric of black $p$-brane solution is:\\
\begin{equation}
ds^2 \ = \ H^{\frac{p+1}{8}}( r ) \left( H^{-1}(r) [ -f(r)dt^2 + {dy_1}^2 + \dots + {dy_p}^2 ] +
f^{-1}(r)dr^2 + r^2 d\Omega_{d+1}^2 \right) \ \ . \label{pmet}\\
\end{equation}
where 
\begin{align}
 H(r) \ &= \ 1 + \dfrac{R^d}{r^d} \ \ , \ \ 
 f(r) \ = \ 1 \ - \dfrac{\mu^d}{r^d} \ \ ,  \ \ 
 R^d \ = \ \mu^d \sinh^2 \gamma \ \  , \\
  d \ &= \  7 - p \ \ , \\  
 \mu \ &\text{:  the horizon radius } \ \ .
\end{align}
The extremal limit corresponds to 
\begin{align}
 \mu \to & 0 \ \ , \\
 \gamma \to & \infty \ \ , \\
 \mu^d \sinh 2\gamma \ & \text{: fixed} \ \ .
\end{align}
The total energy $E$, the charge per unit volume $q$ of the black $p$-brane  are 
\cite{Bp}\cite{adm}\cite{bbd}\cite{spb} , 
\begin{align}
E &= \dfrac{ \omega_{d+1} }{ 2  \kappa^2 } \mu^d  V ( d + 1 \ + \ d \sinh^2 \gamma ) \ ,\\
q &= \dfrac{ \omega_{d+1} }{ 2 \sqrt{2 }  \kappa } d \mu^d \sinh 2\gamma \ ,
\end{align}
where $\omega_{d+1}$ is  volume of a $d+1$-dimensional unit sphere,
$V$ is  volume of the torus which the brane wrapped. \\  
The Bekenstein-Hawking entropy $S$ and the temperature of 
the black $p$-brane are
\begin{align}
 S \ &= \ \dfrac{ 2 \pi \omega_{d+1} }{ \kappa^2 } \mu^{d+1} V H^{1/2} (\mu) \
= \ \dfrac{ 2 \pi \omega_{d+1} }{ \kappa^2 } \mu^{d+1} V \cosh \gamma \ \ ,\\
 T &= \dfrac{d}{4 \pi \mu \cosh \gamma } \ \ . \label{ptempe}
\end{align}  

The absorption probability of black 3-brane for $l$ th partial wave of
dilaton to the lowest order in the radiation frequency is \cite{abs8} 
\begin{equation}
P^{(l)}= \dfrac{ 2^{ -3l -3} \ \pi^2 \  \Gamma(1+l/4) }{ (l+2)!^2 \ \Gamma(1/2+l/4)^2 }
{(\omega \mu)}^{2l+5} \cosh \gamma  \ \ ,
\label{absproba_d3_sugra}
\end{equation}
the absorption cross-section  ( the greybody factor ) of the black 3-brane is written \cite{abs7}
\begin{equation}
{\sigma^{(l)}} = \dfrac{8\pi^2}{3 \omega^5} (l+1)(l+2)^2(l+3)  P^{(l)} \ \ .
\label{greybody_d3_sugra}
\end{equation}

\subsection{Thermodynamical models for black $p$-brane \label{past_models}}

In this subsection, we review some of thermodynamical models for non-extremal black $p$-brane
proposed in the past \cite{B3} \cite{Bp} \cite{DGK}.  In addition, we argue about relation
between those models and this paper's model.

\subsubsection{ Near extremal $p$-brane }

At the beginning, we review thermodynamical model of near extremal black 3-brane \cite{B3}.
`Near extremal' corresponds to the region of $\dfrac{\mu^4}{R^4} << 1 $ in \eqref{pmet}.
In this region, the near horizon geometry of the near extremal black 3-brane is
\begin{align}
ds^2 \  = & \ -\dfrac{r^2}{R^2}\left(1- \dfrac{ \mu^4}{  r^4 } \right) dt^2 \ + \ 
\dfrac{r^2}{R^2} \sum_{i=1}^{3} d {x_i}^2 \   \notag \\
&+  \dfrac{R^2}{r^2} \dfrac{1}{ \left( 1- \dfrac{ \mu^4 }{  r^4 } \right)} dr^2 \ + \  
R^2 \ \ d { \Omega^2_{5} } \ \ \ .
\label{blackbrane3_2}
\end{align}
The entropy is determined by the area of the horizon and the temperature is 
surface gravity on the horizon. Thus the near horizon geometry determines $S$ and $T$.
The entropy and the temperature are written
\begin{align}
S \ &= \ \dfrac{2 \pi^4}{\kappa^2} V R^2 \mu^3 \ \ , \label{neent_1}\\
T \ &= \ \dfrac{\mu}{ \pi R^2 }  \label{netemp_1} \ \ .  
\end{align}
The total energy $E$ can be written as \cite{B3}
\begin{equation}
 E \ = \ M_0 \ + \ \delta M \label{tenergy_ne}
\end{equation}
where $M_0$ is the mass of extremal 3-brane and $\delta M$ is small mass added to $M_0$.
$\delta M$ is written as 
\begin{equation}
 \delta M \ = \ \dfrac{ 3 \pi^3}{ 2 \kappa^2 } V \mu^4 \label{addenergy_ne}
\end{equation}
$S$ and $\delta M$ can be expressed by $R$ and $T$,
\begin{align}
S \ &= \ \dfrac{2 \pi^7}{\kappa^2} R^8 V  T^3 \ \ , \label{neent_2}\\
\delta M  \ &= \ \dfrac{3 \pi^7}{ 2 \kappa^2  }  R^8 V T ^4 \label{netemp_2} \ \ .  
\end{align}
Since $\dfrac{\delta M}{M_0} << 1 $, $R$ in \eqref{neent_2} \eqref{netemp_2} can be written
\begin{equation}
R^4 \ = \ \dfrac{ \sqrt{2} \kappa }{ 4 \pi^3 } q \ \ ,
\label{lr4_ne}
\end{equation}
where $q$ is the charge of the near extremal 3-brane.
\cite{B3} proposed a D3-brane model for near extremal 3-brane. The number of the D3-brane $N$ is \cite{Bp}
\begin{equation}
N \ = \ \dfrac{1}{ \sqrt{2 \pi} } \  q \ \ ,
\label{N_ne3}
\end{equation}
therefore $R^4$ is expressed by $N$ as
\begin{equation}
R^4 \ = \ \dfrac{\kappa}{ 2 \pi^{\frac{5}{2}} } N \ \ .
\label{NR_ne3}
\end{equation}
Thus, $S$ and $\delta M$ are written
\begin{align}
S \ &= \ \dfrac{ 2 \pi^7 }{ \kappa^2 } R^8 V T^3 \ = \ \dfrac{ \pi^2}{2} N^2 V T^3 
\label{neent_3} \ \ , \ \\
\delta M \ &= \ \dfrac{ 3 \pi^7 }{ 2 \kappa^2 } R^8 V T^4 \ = \ \dfrac{3 \pi^2}{ 8 } N^2 V T^4
\label{neenrgy_3} \ \ , \ 
\end{align}
which corresponds to massless open string gas on the D3-branes (however, the numerical factor differs by 
$\dfrac{3}{4}$ from free case) . 

Note that the parameter $R^4$ in the metric \eqref{blackbrane3_2} 
corresponds to $N$, in the above argument.  When $R$ in the near horizon geometry 
\eqref{blackbrane3_2} changes to $R'$, 
\begin{align}
ds^2 \  = & \ -\dfrac{r^2}{R'^2}\left(1- \dfrac{ \mu^4}{  r^4 } \right) dt^2 \ + \ 
\dfrac{r^2}{R'^2} \sum_{i=1}^{3} d {x_i}^2 \   \notag \\
&+  \dfrac{R'^2}{r^2} \dfrac{1}{ \left( 1- \dfrac{ \mu^4 }{  r^4 } \right)} dr^2 \ + \  
R'^2 \ \ d { \Omega^2_{5} } \ \ \ ,
\label{blackbrane3_2e}
\end{align}
the entropy is written
\begin{equation}
S \ = \ \dfrac{ 2 \pi^7 }{ \kappa^2 } {R'}^8 V T^3 \ = \ \dfrac{ \pi^2}{2} {N'}^2 V T^3 
\label{neent_3prime} \ \ , \ 
\end{equation}
which means the number of degrees of freedom of the open string $N^2$ is replaced by $N'^2$.
We will show in the next section that the near horizon geometry of the black 3-brane far from extremality is written 
in similar metric, and the entropy and the energy can be obtained by replacing the number of degree of freedom.

The above argument can easily be extended to general $p$-brane. The near horizon geometry of the near extremal 
black $p$-brane is 
\begin{align}
ds^2 \ = & \ - \dfrac{ r^{d(1-\alpha)}  }{ R^{d(1-\alpha)} } \left(1- \dfrac{ \mu^d }{  r^d } \right) \ dt^2 \ + \ 
\dfrac{  r^{d(1-\alpha)}  }{  R^{d(1-\alpha)}  } \sum_{i=1}^{p} d {x_i}^2 \   \notag \\
&+  \dfrac{ R^{ \alpha d}  }{ r^{ \alpha d }  \left(1- \dfrac{ \mu^d }{  r^d } \right)} dr^2 \ + \  
R^{ \alpha d} r^{2-\alpha d } d { \Omega^2_{d+1} }  \ \ ,
\label{nex_rev_1}
\end{align}
where $\alpha = \dfrac{ p+1}{8}$, and the entropy and the temperature are
\begin{align}
S \ &= \ \dfrac{2 \pi \omega_{d+1} }{ \kappa^2} R^{ \frac{d}{2} } \mu^{ \frac{d}{2} + 1 } V
\ = \ \dfrac{2 \pi \omega_{d+1} }{ \kappa^2 } \left( \dfrac{4 \pi}{d} \right)^{ \frac{\lambda}{1 - \lambda} }
R^{ \frac{ d }{ 2(1-\lambda)} } V T^{ \frac{\lambda}{ 1 - \lambda} } 
\label{S_nep1} \ \ , \ \\
\delta M \ &= \ \dfrac{ \omega_{d+1} }{ 2 \kappa^2 } V \left( \dfrac{d}{2} + 1 \right) \mu^d  \ = \ 
\dfrac{ \omega_{d+1} }{ 2 \kappa^2 }  \left( \dfrac{d}{2} + 1 \right)
\left( \dfrac{4 \pi}{d} \right)^{ \frac{1}{1 - \lambda} } R^{ \frac{d}{2(1-\lambda)} } V T^{ \frac{1}{1-\lambda} }
\label{T_nep1} \ \ , \ 
\end{align}
where $\lambda = \dfrac{8-p}{7-p} - \dfrac{1}{2} $ . 
Those are expressed by the number of D$p$-branes $N$
\begin{equation}
N \ = \ \dfrac{1}{ \sqrt{ 2 \pi } } \dfrac{1}{ ( 2 \pi l_s )^{3-p} } \ q \ \ ,
\label{dpq_ne}
\end{equation}
as
\begin{align}
S \  &=  \ \dfrac{1}{\lambda} C_p N^{\frac{1}{2(1-\lambda)}} T^{\frac{\lambda}{1-\lambda}} \ \
\label{dps_ne1} , \\
\delta M \ &= \ C_p N^{\frac{1}{2(1-\lambda)}}  T^{\frac{1}{1-\lambda}}  \ \ , 
\label{dpdm_ne1}
\end{align}
as shown in \cite{Bp}, where $C_p$ is a constant.
$R^d$ is proportional to $N$ as in 3-brane case. When $R$ in 
\eqref{nex_rev_1} changes, the $N$ in \eqref{dps_ne1} \eqref{dpdm_ne1}
changes.

\subsubsection{Far from extremality}

For the black 3-brane far from extremality, \cite{DGK} proposed the thermodynamical model
\begin{equation}
E \ = \ (N + \bar{N} ) \tau_3 V \ + \ \dfrac{ 3 \pi^2 }{8} N^2 V T^4 \ + \  \dfrac{ 3 \pi^2 }{8} \bar{N}^2 V \bar{T}^4
\label{dgk_energy} \ \ , 
\end{equation}
where $\bar{N}$ is the number of anti D3-branes and $\bar{T}$ is the temperature of the anti D3-branes.
Note that the coefficient $\dfrac{ 3 \pi^2 }{8}$ already includes the $\dfrac{3}{4}$ 
numerical factor of the near extremal case \eqref{neenrgy_3}. 
The entropy is 
\begin{equation}
S \  = \ \dfrac{\pi^2}{2} N^2 V T^4 \ + \ \dfrac{\pi^2}{2} \bar{N}^2 V \bar{T}^4 \ \ .
\label{dgk_entr}
\end{equation}
The correct entropy up to numerical factor ( $2^{\frac{3}{4}}$ ) is reproduced when maximizing  $S$ by $N$ and $\bar{N}$.
Note that the above argument contains two kinds of additional numerical factor. 
The one is the $\dfrac{3}{4}$ factor built in the D3-brane model \eqref{dgk_energy}, another is the discrepancy emerged after the maximization
of the entropy, $2^{\frac{3}{4}}$ \cite{DGK}.

Similar argument can be applied for general $p$-brane \cite{SP} \cite{BL}. Those models also yield discrepancy 
even though they already include numerical factor of the near extremal region.

\subsubsection{Relation to this paper's model}

In the above model for the black $p$-brane far from extremality \cite{DGK} \cite{SP} \cite{BL}, the total
energy after maximizing  the entropy is 
\begin{equation}
E \ = \ (N + \bar{N} ) \tau_p V \ + \ 4 \lambda \sqrt{ N \bar{N} } \tau_p V
\label{dgkp_energymax} \ \ .
\end{equation}
where $N$, $\bar{N}$ are written by the supergravity parameters as
\begin{align}
N \ - \ \bar{N}  \ &= \ \dfrac{ \omega_{d+1} d  }{ 4 \kappa^2 \tau_p } \mu^d \sinh 2 \gamma
\ =  \ \dfrac{ \omega_{d+1} d  }{ 2 \kappa^2 \tau_p } R^{d/2} ( R^d + \mu^d )^{1/2} \ , \\
\sqrt{N \bar{N} } \ &= \  \dfrac{ \omega_{d+1} d  }{ 8 \kappa^2 \tau_p } \mu^d
\label{dgkp_add} \ \ .
\end{align}
The second term of \eqref{dgkp_energymax} corresponds to the energy of gas on the branes and the antibranes.
If we rewrite \eqref{dgkp_energymax} by $Q=N-\bar{N}$,
\begin{equation}
E \ = \ \sqrt{ Q^2 +  ( 2 \sqrt{ N \bar{N} } )^2 } \ \tau_p V \ + \  4 \lambda \sqrt{ N \bar{N} } \ \tau_p V
\label{dgkp_energymax2} \ \ .
\end{equation}
$E$ and $Q$ are fixed when maximizing the entropy. We have now one parameter $\sqrt{N \bar{N}}$ instead of 
$N$ and $\bar{N}$. The $\sqrt{N \bar{N}}$ determines the degree of non-extremality. If we denote
\begin{equation}
n \ = \ 2 \sqrt{N \bar{N}}
\label{n_nnbar1} \ \ ,
\end{equation}
the energy is 
\begin{align}
E \ &= \ \sqrt{ Q^2 +  n^2 } \ \tau_p V \ + \  2 \lambda n \ \tau_p V \ \ \label{dgkp_energymax_n0}, \\
\ &= \sqrt{ Q^2 +  n^2 } \ \tau_p V \ + \  C_p ( \sqrt{ Q^2 +  n^2 }  + n )^{ \frac{1}{2(1-\lambda)} } V 
T^{\frac{1}{1-\lambda}}  
\label{dgkp_energymax_n} \ \ .
\end{align}
where $C_p$ is a constant. The above \eqref{dgkp_energymax_n0}  \eqref{dgkp_energymax_n} is the energy {\it after maximizing the entropy }.

Our model is as follows. The total energy {\it before maximizing the entropy} is
\begin{equation}
E \ = \ \sqrt{ Q^2 +  n^2 } \ \tau_p V \ + \  C_p ( \sqrt{ Q^2 +  n^2 }  + n )^{ \frac{1}{2(1-\lambda)} } V 
T^{\frac{1}{1-\lambda}}  
\label{our_energy1} \ \ ,
\end{equation}
where $n$ is a free parameter, and the correct entropy and temperature are obtained when maximizing the entropy by $n$. This model smoothly reduces
to the near extremal model \cite{B3} \cite{Bp} when $\dfrac{n}{N} << 1$.  TThe constant $C_p$ is determined by comparing with the near extremal models \cite{B3} \cite{Bp} 
in the near extremal region. Once we include the numerical factor in the near extremal model as $C_p$, our model 
does not yield any other discrepancies, and correctly reproduces entropy and temperature through 
all the non-extremal region. 

%%%%%%%%%%%%%%%%%%%%%%%%%%%%%%%%%%%%%%%%%%%%%
%
%
%      honro
%
%
%%%%%%%%%%%%%%%%%%%%%%%%%%%%%%%%%%%%%%%%%%%%

\section{Thermodynamics  of non-extremal black $p$-brane \label{honron}}

\subsection{An ansatz of  thermodynamical model   } 

First, we introduce an ansatz of microcanonical thermodynamical model for 
non-extremal black $p$-brane as follows. In the next subsection,
we argue about the appropriateness of this ansatz.\\
\mbox{}\\
\mbox{}
\underline{Ansatz}\\
----------------------------------------------------------------------------------------\\
Partition function $Z$ for non-extremal black $p$-brane thermodynamics:
\begin{equation}
\ln Z \ = \ -\dfrac{ 1}{ T } \left( \sqrt{ N^2 + n^2 }  - N \right) \tau_p V  \ +  \
\dfrac{ 1-\lambda}{ \lambda }  f( N, n ) V T^{ \frac{ \lambda} { 1- \lambda } } \ \ ,
\label{parti_model }
\end{equation}
where 
\begin{align}
f(N, n ) \ &= \ C_p \left( \sqrt{ N^2 + n^2 } + n  \right)^{\frac{1}{ 2(1-\lambda) } } \ \ \ , \\
C_p & = 2^{ \frac{2(15-2p)}{5-p} } d^{ - \frac{2(8-p)}{5-p} }  \lambda \ \pi^{ \frac{ 2(7-p) }{ 5-p } } 
\omega_{d+1}^{- \frac{2}{5-p}  } \kappa^{ \frac{4}{5-p} } \tau_p^{ \frac{d}{5-p} } \ \ \ , \\
\lambda &= \dfrac{8-p}{7-p} - \dfrac{1}{2} \ \ \ , \\
N &: \text{ charge } ,\ \ \ V : \text{ The volume of the brane  } ,\notag
\label{ fdef}
\end{align}
and  $n$ is determined so that the entropy is maximized under  fixed energy and charge.
The resulted  $n$ is $n = \dfrac{ \omega_{d+1} d }{ 4 \kappa^2 \tau_p } \mu^d$ .\\

Total energy of the black $p$-brane $E_{total}$ is
\begin{equation}
E_{total} = E_{thermal} + N \tau_p V \ \ ,
\end{equation}
where $N \tau_p V $ is the mass of the black $p$-brane in the extremal limit (i.e. zero temperature brane).

----------------------------------------------------------------------------------------\\
\mbox{}\\
\mbox{}
\mbox{}\\
\mbox{}
From the ansatz \eqref{parti_model } we get (thermal) energy and free energy and entropy:\\
\mbox{}\\
\mbox{}

(i)   Energy:\\
\begin{align}
E_{thermal} \ & = \ T^2  \dfrac{\partial }{\partial T } \ln Z \notag \\
&= \ \ \left( \sqrt{ N^2 + n^2 } - N  \right) \tau_p V \ + \ f (N, n) V T^{ \frac{1}{1-\lambda} }
\label{thermal_energy_model }
\end{align}

(ii) Free energy:\\
\begin{align}
F \ &= \ - T \ln Z  \notag\\
& = \ \left( \sqrt{ N^2 + n^2 } - N  \right) \tau_p V \ - \ \dfrac{1-\lambda}{\lambda} f(N,n) V T^{ \frac{1}{1-\lambda} }
\label{free_energy_model }
\end{align}

(iii) Entropy:\\
\begin{align}
S \ &= \ \dfrac{E_{thermal} - F }{ T}  \notag \\
     &= \ \dfrac{1}{\lambda } f(N,n) V T^{ \frac{\lambda}{1-\lambda} }
\label{entropy_model }
\end{align}
\mbox{}\\
The $n$ and temperature $T$ are determined by the condition of
maximum entropy, as follows.  \\

Rewrite the  entropy as\\
\begin{align}
S \ &= \ \dfrac{1}{\lambda} f V T^{ \frac{\lambda}{1-\lambda} } \notag \\
& = \ \dfrac{1}{\lambda}  f^{1-\lambda} V^{1-\lambda }  
( E_{thermal} - \sqrt{ N^2 + n^2 }\tau_p V + N\tau_p V )^\lambda   \ \ \ .
\label{entropy_max1 }
\end{align}
The condition 
\begin{equation}
\dfrac{ \partial  S }{ \partial n } = 0 \ \ ,
\label{max_entropy1 }
\end{equation}
at fixed $ E_{thermal} $ ( or $E_{total}$ ) and fixed charge $N$ yields the following equation
\begin{equation}
( E_{thermal} - \sqrt{ N^2 + n^2 }\tau_p V + N\tau_p V ) = 2 \lambda n \tau_p V \ \ .
\label{max_bibun }
\end{equation}
Then the total energy $E_{total}$ is written as 
\begin{equation}
E_{total }  =  \sqrt{ N^2 + n^2 }\tau_p V + 2 \lambda n \tau_p V  \ \ \ . 
\label{max_bibun2}
\end{equation}
From above \eqref{max_bibun2} and the given total energy and charge, we get $n$.
Also, we get the temperature as
\begin{equation}
T = \left( \dfrac{ 2 \lambda n \tau_p }{ f(N,n ) }  \right)^{ 1 - \lambda } \ \ \ .
\label{temp_max_ent}
\end{equation}
The second order derivative of the entropy is
\begin{equation}
\dfrac{ \partial^2  S }{ \partial n^2 } = - \lambda \sqrt{ N^2 + n^2 } \ - \ 2 \lambda^2 (N^2 + n^2 ) \ < \ 0 \ \ ,
\label{entropy_nikaibibun }
\end{equation}
and the first order derivative vanishes at the one point, thus the above solution
is the solution at the maximum entropy.\\
\mbox{}\\
\mbox{}
The model  has the following characteristics:

\begin{enumerate}
\renewcommand{\labelenumi}{(\roman{enumi})}
\item For a given total energy and charge,  the ansatz yields the correct temperature and entropy, 
which agree with the supergravity result.
\item The thermal energy $E_{thermal}$ vanishes at extremal limit.
\item The model reduces to near extremal black $p$-brane thermodynamical model \cite{B3} \cite{Bp} 
at $\dfrac{n}{N} \to 0 $ limit.
\item The second term of \eqref{thermal_energy_model }  can be obtained 
from the thermodynamical model  of {\it near extremal} black $p$-brane 
by replacing the "freedom of the gas" by $f(N,n)$.\label{ne_model_char}
\item When we substitute the "freedom of the gas" $f(N,n)$ and the temperature to the 
absorption probability of {\it near extremal} black 3-brane, we get the correct absorption probability
and greybody factor of  the {\it non-extremal} black 3-brane. \label{absorption_char}
\end{enumerate}
We explain   \ref{ne_model_char} and \ref{absorption_char} in the following.\\
\mbox{}\\
\mbox{}
\mbox{}\\
\mbox{}
\underline {Characteristic  \ref{ne_model_char} } \\
\mbox{}\\
\mbox{}
The thermodynamical model of the near extremal black 3-brane  is \cite{B3}
\begin{equation}
E_{thermal} = C_3 N^2 V T^4 , 
\end{equation}
where $C_3$ is constant ( $ 3 \pi^2 / 8 $) . The $N$ is the number of coincident D3-branes,
and  $N^2$  is the number of degrees of freedom of the open string. Replacing the $C_3 N^2$ by $f(N,n)$ 
\begin{equation}
C_3 N^2  \ \ \to \ \  C_3 ( \sqrt{N^2+n^2} + n )^2
\end{equation}
yields the second term of the  \eqref{thermal_energy_model }. 
For $p \neq 3$, the replacement
\begin{equation}
C_p N^{\frac{1}{2(1-\lambda)} } \ \ \to \ \  C_p ( \sqrt{N^2+n^2} + n )^{\frac{1}{2(1-\lambda)}} \ 
\end{equation}
in the near extremal $p$-brane model \cite{Bp} yields the second term of \eqref{thermal_energy_model } \\
\mbox{}\\
\mbox{}
\underline {Characteristic \ref{absorption_char} } \\
\mbox{}\\
\mbox{}
In the following, we show that the absorption probability (and the greybody factor)
 of  {\it non-extremal } black 3-brane can be  obtained by replacing ``the number of the degrees of freedom" 
and temperature of the  {\it near extremal}  result.
 
 The absorption probability of near extremal black 3-brane for the $l$th partial wave  calculated 
by supergravity theory is \cite{abs2} \cite{abs8}
\begin{equation}
 P^{(l)} = \dfrac{ 2^{ -2l -3} \ \pi \  \Gamma(1+l/4)^4 }{ (l+2)!^2 \ \Gamma(1+l/2)^2 }
\omega^{2l+5} \mu^{2l+3} R^2  \ \ .
\end{equation}
In the near extremal region, the parameter $R$ and $\mu$ in supergravity can be written by $N$ and $T$ as \cite{abs8}
\begin{equation}
 R^4 = \dfrac{ \kappa N }{ 2 \pi^{5/2 } }, \ \ \ \ \ 
\mu = T\sqrt{\dfrac{ \kappa N }{ 2 \pi^{1/2 } } } \ \ .
\end{equation}
Rewriting the absorption probability by $N$ and $T$,
\begin{equation}
 P^{(l)} = \dfrac{ 2^{ -2l -3} \ \pi \  \Gamma(1+l/4)^4 }{ (l+2)!^2 \ \Gamma(1+l/2)^2 }
\omega^{2l+5} N^{l+2} T^{2l+3} \kappa^{l+2} \left( \dfrac{1}{2} \right)^{l+2} \left( \dfrac{1}{\pi} \right)^{(l+4)/2} 
\label{apbyN} \ \ .
\end{equation}
 Replacing \ $N$ \ by \ $\sqrt{N^2+n^2} + n$ \ 
and substituting the correct $T$, and rewriting those by supergravity parameters, we get
\begin{equation}
 P^{(l)} = \dfrac{ 2^{ -3l -3} \ \pi^2 \  \Gamma(1+l/4) }{ (l+2)!^2 \ \Gamma(1/2+l/4)^2 }
{(\omega \mu)}^{2l+5} \cosh \gamma  \ \ .
\label{apbyNn}
\end{equation}
This result agrees with the result from the supergravity \eqref{absproba_d3_sugra}. 
The greybody factor is calculated as \eqref{greybody_d3_sugra}, thus we reproduced  
greybody factor of black 3-brane by this "replacing".

%%%%%%%%%%%%%%%%%%%%%%%%%%%%%%% near extremal and... %%%%%%%%%%%%%%%%%%%%%%%%%%

\subsection{Near extremal and non-extremal geometry} 

In this section, we show that the non-extremal black $p$-brane has a 
relation to near extremal black $p$-brane geometry.  In the vicinity of the 
horizon, the non-extremal $p$-brane (far from extremality) has similar geometry
with near extremal $p$-brane, and this leads to the temperature and 
the entropy described by the similar form.  We also show that this geometrical similarity
has a relation to various properties of our thermodynamical model.

\subsubsection{The vicinity of the horizon geometry }

A non-extremal black $p$-brane  (arbitrary far from extremality) with parameter $R_0$ and $\mu_0$ is \\
\begin{align}
ds^2 \ = & \ -\dfrac{ \left(1- \dfrac{ \mu_0^d }{  r^d } \right) }{ \left( 1 + \dfrac{ R_0^d }{ r^d } \right) ^{1-\alpha}} dt^2 \ + \ 
\dfrac{ 1 }{ \left( 1 + \dfrac{ R_0^d }{ r^d } \right) ^{1-\alpha} } \sum_{i=1}^{p} d {x_i}^2 \   \notag \\
&+  \dfrac{\left( 1 + \dfrac{ R_0^d }{ r^d } \right)^\alpha }{ \left( 1- \dfrac{ \mu_0^d }{  r^d } \right)} dr^2 \ + \  
\left( 1 + \dfrac{ R_0^d }{ r^d } \right) ^\alpha r^2 d { \Omega^2_{d+1} } \ \ \ ,
\label{black_taiou}
\end{align}
where
\begin{equation}
\alpha = \dfrac{p+1 }{ 8}  \ \ \ .
\label{ alpha_honron}
\end{equation}
On the other hand, the near horizon limit of {\it near extremal} 
black $p$-brane parameterized by $R$ and $\mu$ is \\
\begin{align}
ds^2 \ = & \ - \dfrac{ r^{d(1-\alpha)}  }{ R^{d(1-\alpha)} } \left(1- \dfrac{ \mu^d }{  r^d } \right) \ dt^2 \ + \ 
\dfrac{  r^{d(1-\alpha)}  }{  R^{d(1-\alpha)}  } \sum_{i=1}^{p} d {x_i}^2 \   \notag \\
&+  \dfrac{ R^{ \alpha d } }{ r^{ \alpha d }  \left(1- \dfrac{ \mu^d }{  r^d } \right)} dr^2 \ + \  
R^{ \alpha d} r^{2-\alpha d } d { \Omega^2_{d+1} } \ .
\label{ne_taiou}
\end{align}
When we change the parameters $R$ and $\mu$  as 
\begin{align}
R^d \ &= \ {R_0}^d + {\mu_0}^d \ \ , \\
\mu &= \mu_0 \ \ ,
\label{Rmutaiou1 }
\end{align}
\eqref{ne_taiou} is
\begin{align}
ds^2 \ = & \ - \dfrac{ r^{d(1-\alpha)}  }{ ({R_0}^d + {\mu_0}^d)^{(1-\alpha)} } \left(1- \dfrac{ \mu_0^d }{  r^d } \right) \ dt^2 \ + \ 
\dfrac{  r^{d(1-\alpha)}  }{  ({R_0}^d + {\mu_0}^d)^{(1-\alpha)}  } \sum_{i=1}^{p} d {x_i}^2 \   \notag \\
&+  \dfrac{ ({R_0}^d + {\mu_0}^d)^{ \alpha} }{ r^{ \alpha d }  \left(1- \dfrac{ \mu_0^d }{  r^d } \right)} dr^2 \ + \  
({R_0}^d + {\mu_0}^d)^{ \alpha } r^{2-\alpha d } d { \Omega^2_{d+1} } \ .
\label{ne_taiou2}
\end{align}
First, the non-extremal $p$-brane \eqref{black_taiou} and 
the parameter changed near extremal $p$-brane \eqref{ne_taiou2} are equal at the horizon ($r=\mu_0$).

Next, we consider the vicinity of the horizon. When we expand the metric for 
the time direction  of the  non-extremal $p$-brane \eqref{black_taiou} (we denote it as   $g_{tt}(r)$ ) in the vicinity 
of  the horizon,
\begin{equation}
g_{tt}( \mu_0 + \epsilon ) = g_{tt}(\mu_0) - \dfrac{ \mu_0^{d(1-\alpha) - 1 } d } {(R_0^d+\mu_0^d)^{1-\alpha} } \epsilon   \ \ \ .
\label{ gttkinbou }
\end{equation}
On the other hand, when we expand the metric for 
the time direction of the  parameter changed near extremal $p$-brane \eqref{ne_taiou2}
(we denote it as  $h_{tt}(r)$) in the vicinity of  the horizon,
\begin{equation}
h_{tt}( \mu_0 + \epsilon ) = h_{tt}(\mu_0) - \dfrac{ \mu_0^{d(1-\alpha) - 1 } d } {(R_0^d+\mu_0^d)^{1-\alpha} } \epsilon  \ \ \ .
\label{ httkinbou }
\end{equation}
The both agree.  The same agreement holds for the  $r$ direction.  
Both metrics for $r$ direction: $g_{rr}$ and $h_{rr}$ also agree 
in the vicinity of  the horizon.

%%%%%%%%%%%%%%%%%%%%%%%%%%%%%%%%%%%%%%%%%%%%%%%
\subsubsection{The entropy and the temperature  \label{match_enttemp} }

Consider the following two types of geometry: \\ 
\mbox{}\\
\mbox{}
(i) Geometry-A\\
\hspace{3ex} Geometry \eqref{ne_taiou2} in near horizon region,
and asymptotically flat. \\
\mbox{}\\
\mbox{}
(ii) Geometry-B\\
\hspace{3ex} Non-extremal black $p$-brane background \eqref{black_taiou}.  \\
\mbox{}\\
\mbox{}
In the following, we show that the entropy and the temperature of 
the both geometry agree. \\
\mbox{}\\
\mbox{}\\
\mbox{}
\underline{Entropy}\\

As we mentioned in the previous subsection, the metric of the geometry A and 
the geometry B agree at the horizon $r=\mu_0$.  Thus their entropies agree.\\
\mbox{}\\
\mbox{}\\
\mbox{}
\underline{Temperature}\\

Defining $u$ as
\begin{equation}
 u \ = \ 2 \ \left(\dfrac{(R_0^d+ \mu_0^d)^\alpha }{\mu_0^{\alpha d - 1 } d }\right)^{1/2} ( r- \mu_0 )^{1/2} \ \ \ ,
\label{ u_teigi_taiou}
\end{equation}
both metrics in the vicinity of the horizon can be written in
Euclidean form as 
\begin{equation}
ds^2 \ \sim \ u^2 \left(\dfrac{dt_E}{2 \dfrac{(R_0^d+ \mu_0^d)^{1/2}}{\mu_0^{d/2 - 1 } d}}\right)^2
\ +  \ du^2 
\ +  \ ( \text{other terms} ) \ \ \ .
\label{kinboum_metric2_taiou }
\end{equation}
By comparing above with the polar coordinates of 2-dimensional plane 
\begin{equation}
ds^2 \ =\ u^2 d\theta^2 \ + \ du^2 \ , \ \  \ 0 \le \theta \le 2 \pi \ \ ,
\label{kyokuzahyou }
\end{equation}
the period of the compactified $t_E$(which is $\dfrac{1}{T}$) must be
\begin{equation}
\dfrac{1}{T} \ = \ \dfrac{ 4\pi (R_0^d+ \mu_0^d)^{1/2} }{ \mu_0^{d/2 - 1 } d } \ \ \ ,
\label{tbunno1 }
\end{equation}
so that conical singularity in $t-r$ plane vanishes.
As above, the temperature of the two geometries agree because their $tt-$ and $rr-$
components are equal in the vicinity of the horizon.\\
\mbox{}\\
\mbox{}\\
\mbox{}
\underline{Near extremal and far from extremality}\\

As we mentioned in section \ref{past_models}, near extremal region corresponds to
the region of $\dfrac{\mu}{R} << 1$. In this region,
\begin{align}
S \  &\sim \ \dfrac{2 \pi \omega_{d+1} }{ \kappa^2 } \mu_0^{\frac{d}{2}+1} V \sqrt{ R_0^d } \ \ , \label{net_s_1} \\
T \ &\sim \ \dfrac{ \mu_0^{\frac{d}{2}-1} d }{ 4 \pi \sqrt{R_0^d } } \ \ . \label{net_t_1} 
\end{align}
On the other hand, $\dfrac{\mu}{R}$ can not be neglected in the region far from extremality. In this region,
\begin{align}
S \  &\sim \ \dfrac{2 \pi \omega_{d+1} }{ \kappa^2 } \mu_0^{\frac{d}{2}+1} V \sqrt{ R_0^d + \mu_0^d} \ \ , \label{net_s_2} \\
T \ &\sim \ \dfrac{ \mu_0^{\frac{d}{2}-1} d }{ 4 \pi \sqrt{R_0^d + \mu_0^d} } \ \ . \label{net_t_2} 
\end{align}
One can see from \eqref{net_s_1} \eqref{net_t_1} and \eqref{net_s_2} \eqref{net_t_2} that 
$S$ and $T$ of ``far from extremality" region are obtained by replacing $R_0^d \to R_0^d + \mu_0^d$
in the near extremal $S$ and $T$. The geometric similarity we showed above corresponds to this `replacement'.
As we have shown in section \ref{past_models}, $R_0^d$ of near extremal $p$-brane corresponds to the number of D$p$-branes $N$.
Thus above replacement corresponds to the replacement of $N$ in the near extremal thermodynamical model.
Explicitly, $R_0^d \to R_0^d + \mu_0^d$ corresponds to 
\begin{equation}
N \ \to \ \sqrt{N^2+n^2} + n \ \ \ ,
\label{Ntonnnn1 }
\end{equation}
and this leads to the second term of total energy \eqref{thermal_energy_model }
\begin{equation}
E_{thermal} \  = \  \left( \sqrt{ N^2 + n^2 } - N  \right) \tau_p V \ + \ f (N, n) V T^{ \frac{1}{1-\lambda} }
\label{e_thermal2 }
\end{equation}
of our model, where $n$ is determined by the maximization of entropy in our model as 
\eqref{entropy_max1 } to \eqref{temp_max_ent}.

%%%%%%%%%%%%%%%%%%%%%%%%%%%%%%%%%%%

\subsubsection{Match of absorption probability and greybody factor }

We show below that the equation of the motion of the massless scalar field in 
geometry A 3-brane and geometry B 3-brane is equal in the low energy region. 
This equality results in the agreement of the absorption probability (and the greybody factor).

The equation of the motion of massless scalar field in non-extremal black 3-brane  background (the geometry B ) is
\begin{align}
\partial_r^2 \phi \ & + \ \left(\dfrac{5}{r}+\dfrac{4\mu^4}{r^5 \left( 1 - \dfrac{\mu^4}{r^4}\right) }  \right) \partial_r \phi \notag \\
&\ + \left( \dfrac{ 1+\dfrac{R^4}{r^4}  }{ \left(1 - \dfrac{\mu^4}{r^4}\right)^2 }\omega^2 \ - \ \dfrac{l(l+4) }{ r^2\left(1 - \dfrac{\mu^4}{r^4}\right)  } 
\ - \ \dfrac{1+\dfrac{R^4}{r^4}}{1 - \dfrac{\mu^4}{r^4}} k^2 \right) \phi = 0 \ ,
\label{ p3scalar1}
\end{align}
where\\
\mbox{}\\
\hspace{4ex}$\omega$: energy  ( of the scalar field $\phi$ ) \ \ , \\
\hspace{4ex}$l$: angular momentum \ \ , \\
\hspace{4ex}$k$: momentum in direction to $x_i$ \ \ . \\
\mbox{}\\
We consider the equation of the motion in the low energy region $\dfrac{ \omega }{ T } << 1$ (where $T$ is the 
temperature of the black 3-brane) in the following. \\

The equation of motion of massless scalar field in the geometry A of near horizon region is
\begin{align}
\partial_r^2 \phi \ & + \ \left(\dfrac{5}{r}+\dfrac{4\mu^4}{r^5 \left( 1 - \dfrac{\mu^4}{r^4}\right) }  \right) \partial_r \phi \notag \\
&\ + \left( \dfrac{ \dfrac{R^4+\mu^4}{r^4}  }{ \left(1 - \dfrac{\mu^4}{r^4}\right)^2 }\omega^2 \ - \ \dfrac{l(l+4) }{ r^2\left(1 - \dfrac{\mu^4}{r^4}\right)  } 
\ - \ \dfrac{\dfrac{R^4 + \mu^4}{r^4}}{1 - \dfrac{\mu^4}{r^4}} k^2 \right) \phi = 0 \ \ \ .
\label{ p3scalarA_nh}
\end{align}
At infinity of geometry A, the equation is the same as that in flat spacetime.

The difference of the equation in the geometry A and that in the geometry B is
the terms which contain $R$.  When we consider the massless particle 
propagating  perpendicular to the brane ($k=0$), the difference of the equation in
the geometry A and that in the geometry B is  the $\omega$ term only.\\
\mbox{}\\
\mbox{}\\
\mbox{}
(i) Outer region ($ \mu << r  $ ($ \rho_h << \rho$) )\\

With 
\begin{equation}
\rho = \omega r, \ \  \rho_h = \omega \mu  \ \ \ ,
\label{scalar_rho }
\end{equation}
the equation in the geometry B (black 3-brane) is
\begin{align}
\partial_\rho^2 \phi \ & + \ \dfrac{5 \rho^4 - \rho_h^4}{\rho(\rho^4-\rho_h^4)} \partial_\rho \phi
 - \dfrac{\rho^2 l(l+4)}{\rho^4-\rho_h^4} \phi + \dfrac{\rho^4(\rho^4+(\omega R)^4)}{(\rho^4-\rho_h^4)^2}\phi = 0 
\label{ p3scalar2}
\end{align}
\mbox{}\\
\mbox{}\\
When $\mu < R$, the equation in the outer region of the geometry B(black 3-brane background) is
\begin{equation}
\partial^2_\rho \phi +\dfrac{5}{\rho}\partial_\rho \phi+ \left( 1 + \dfrac{(\omega R)}{\rho^4}
-\dfrac{l(l+4)}{\rho^2} \right)\phi = 0 \ \ \ .
\label{ outer_scalar }
\end{equation}
From the low energy condition
\begin{equation}
\dfrac{(\omega R)^4 }{\rho^2}  <<  \left( \dfrac{ \omega (R^4+\mu^4)^{1/2} }{\mu} \right)^2 
= \left(\dfrac{\omega}{\pi T} \right)^2  << 1 \ \  ,
\label{outer_muT }
\end{equation}
the $R$ term can be ignored.  When ignoring the $R$ term, the equation of the motion
is the same as that in the flat space.  Thus the equation in the geometry B is equal to 
that in the geometry A.

When $\mu  \sim R$ or  $R < \mu$,  the equation in the geometry B is the same as 
the equation in flat spacetime from the condition of the outer region. 
Thus the  equation in the geometry B is equal to that in the geometry A again.

In summary, for every case of $R$ and $\mu$, the equation in the geometry A and that 
in the geometry B are equal in the outer region.\\
\mbox{}\\
\mbox{}\\
\mbox{}
(ii) Inner region ( $\dfrac{\mu}{r} \sim O(1)$ and $\mu \le r$ ) \\

When $\mu < r$, the equation in the  geometry B(black 3-brane) is

\begin{align}
\partial_r^2 \phi \ & + \ \left(\dfrac{5}{r}+\dfrac{4\mu^4}{r^5 \left( 1 - \dfrac{\mu^4}{r^4}\right) }  \right) \partial_r \phi \notag \\
&\ + \left( \dfrac{ 1+\dfrac{R^4}{r^4}  }{ \left(1 - \dfrac{\mu^4}{r^4}\right)^2 }\omega^2 \ - \ \dfrac{l(l+4) }{ r^2\left(1 - \dfrac{\mu^4}{r^4}\right)  } 
\ \right) \phi = 0 \ \ \ .
\label{ p3scalar12}
\end{align}
The $R$ term can be ignored since $\dfrac{ \omega }{ T } << 1$.
The equation in the geometry A is 
\begin{align}
\partial_r^2 \phi \ & + \ \left(\dfrac{5}{r}+\dfrac{4\mu^4}{r^5 \left( 1 - \dfrac{\mu^4}{r^4}\right) }  \right) \partial_r \phi \notag \\
&\ + \left( \dfrac{ \dfrac{R^4+\mu^4}{r^4}  }{ \left(1 - \dfrac{\mu^4}{r^4}\right)^2 }\omega^2 \ - \ \dfrac{l(l+4) }{ r^2\left(1 - \dfrac{\mu^4}{r^4}\right)  } 
\  \right) \phi = 0 \ \ .
\label{ p3scalarA_nh2}
\end{align}
The $R$ term also can be ignored because $\dfrac{ \omega }{ T } << 1$.
Thus the equation in the geometry A and that in the
geometry B is the same, in the region of  $\mu < r$ in the inner region.

In the region of $\mu \sim r$, the $R$ term cannot be ignored in both geometries,
but $g^{rr}$ is the same in  the region of $\mu \sim r$.
The $R$ term in both equations agree since the $R$ term is actually $(1/g^{rr})^2$.
Thus both equations agree in the inner region.\\
\mbox{}\\
\mbox{}
In summary,
\begin{enumerate}
\renewcommand{\labelenumi}{(\roman{enumi})}
\item $\mu << r $ \ \ Both equations are the equation in flat space.
\item $\mu < r $ \ \  Both equations agree by the low energy condition.
\item $\mu \sim r$ \ \ Both equations agree since the $1/g^{rr}$ is the same in the vicinity of the horizon.
\end{enumerate}
Thus the equation in the geometry A and that in the geometry B are equal in
all the region.\\
\mbox{}\\
\mbox{}
The geometry A is a parameter changed near extremal geometry ($R_0^4 \to R_0^4 + \mu_0^4$) in the near horizon region.
Since the scalar field equation in both geometries are the same, the absorption probability
of ``far from extremality" region is obtained by replacing $R_0^4 \to R_0^4 + \mu_0^4$
in the near extremal absorption probability. As we explained, this replacement corresponds to the replacement  
$N \to \sqrt{N^2+n^2} + n$, where $n$ is determined by the maximization of entropy in our model. 
The characteristic \ref{absorption_char} of our thermodynamical model ansatz: \\
\mbox{}\\
\mbox{}
------------------------------------------------------------------------------------------------\\
\mbox{}
When we substitute the "freedom of the gas" $f(N,n)$ and the temperature to the 
{\it near extremal} black 3-brane result of absorption probability, we get the correct absorption probability
and greybody factor of  the {\it non-extremal} black 3-brane. \\
------------------------------------------------------------------------------------------------\\
\mbox{}\\
\mbox{}
caused by the above equality of the equation in the geometry A and  B.
\vspace{3ex}

\subsection{ Appropriateness of the thermodynamical model  \label{maxentro }}

\subsubsection{ Gas term  (after maximization of entropy) }

As we have shown, non-extremal $p$-brane geometry is similar to `parameter changed near extremal 
$p$-brane geometry' in the vicinity of the horizon. This leads to the agreement of entropy, temperature
and absorption probability in both the geometry A and B.
Hence, we can describe thermodynamics of non-extremal $p$-brane by the near extremal $p$-brane model
with replacement $N \to \sqrt{N^2 + n^2} + n$, which results in the second term (gas term) of 
\begin{equation}
E_{thermal} \  = \  \left( \sqrt{ N^2 + n^2 } - N  \right) \tau_p V \ + \ f (N, n) V T^{ \frac{1}{1-\lambda} }
\ \ \ , \label{e_thermal3 }
\end{equation}
where $n$ is determined by the maximization of entropy in our model.

The original explanation of the black hole Hawking radiation \cite{hawk1} is derived by 
considering the behavior of massless scalar field in black hole background. In order to
examine physical appropriateness of our model more strictly, we have to examine the action
of massless scalar field (in this case, dilaton field) in black $p$-brane background.
 We show below that the action 
in the geometry A and the geometry B actually agree in low energy region for $p = 3$.

The Lagrangian of massless scalar field  in non-extremal black 3-brane background (geometry B) is
\begin{align}
 \mathcal{L} &= \sqrt{-g} g^{\mu \nu } \partial_\mu \phi \partial_\nu \phi  \ \ \ , \\
&= r^5 \left( - \dfrac{H}{f} (\partial_t \phi)^2 + f (\partial_r \phi)^2 +
\sum_{i=1}^3 H (\partial_i \phi)^2 + 
\sum_{j=1}^5 \dfrac{1}{ r^2 \Omega_i(\theta_1, \cdots, \theta_5 ) } ( \partial_{\theta_j} \phi)^2 \right) ,
\label{scalar_lag}
\end{align}
where
\begin{align}
H(r) &= 1 + \dfrac{R^4}{r^4}, \ \ \ f(r) = 1 - \dfrac{\mu^4}{r^4}\ \ \ , \\
r^2 \Omega_i \ &: \  \text{ the metric for the angular directions} \ \theta_i.
\label{where_rel_therm }
\end{align}

On the other hand, for massless scalar field in the `parameter changed 
near extremal black 3-brane' background, $H(r) = \dfrac{ R^4 + \mu^4}{ r^4 } $ \ .
Note that $\dfrac{f}{H}$ is the same for both backgrounds at the vicinity of the horizon.

If the $x_i$ direction ( 3 dimensional directions ) are compactified and the size
of the compactified manifold is small enough compared to the energy scale of $\phi$ , 
the $(\partial_i \phi)^2$ term can be dropped from the Lagrangian.
Then the difference is only in the $(\partial_t \phi)^2$ term, and this term
is the same in the vicinity of the horizon for both backgrounds.

Now, consider the low energy limit $\omega \mu \to 0$. The Lagrangian of the massless scalar 
field at the energy scale of $\omega \mu \to 0$ vanishes in almost all the region of 
the space, except near the horizon, where $f(r)$ diverges.  The dominant contribution 
to the Euclidean action is the action near the horizon, thus 
in the low energy limit $\omega \mu \to 0 $, the Euclidean action in the geometry A
and that in the geometry B are approximately the same. The agreement of the Euclidean action means
the same quantum phenomena of the same temperature.  Hence, applying the 
thermodynamical model of the geometry A to the thermodynamics of the geometry B 
is regarded to be valid in this limit.

The limit $\omega \mu \to 0 $ corresponds to blackbody radiation region, 
as one can see that the greybody factor \eqref{greybody_d3_sugra} is non-zero only for $l=0$
where it does not depend on the energy.  Thus our thermodynamical model successfully 
reproduces the blackbody radiation of the non-extremal black 3-brane.

\subsubsection{Total energy and maximization of entropy}

The gas term we explained in the previous subsection requires specific value of
$n$. However, our model does not require $n$ as an input. 

The total energy of our model is
\begin{equation}
E_{thermal} \  = \  \left( \sqrt{ N^2 + n^2 } - N  \right) \tau_p V \ + \ f (N, n) V T^{ \frac{1}{1-\lambda} }
\ \ \ , \label{e_thermal4 }
\end{equation}
where $n$ is a {\it free parameter}. 
When we maximize the entropy by $n$ under fixed $E$ and $N$, the value of $n$ is automatically determined.
The $n$ at the maximum entropy yields correct entropy, temperature and absorption probability.
This property is non-trivial, because free parameters like this generally do not yield 
correct entropy, as discrepancies found  in \cite{DGK} \cite{SP} \cite{BL}.

In addition, our model smoothly reaches to near extremal model \cite{B3} \cite{Bp} when $\dfrac{n}{N} << 1$.
The coefficient in the gas term ( the second term of the energy \eqref{e_thermal4 } ) is determined
by comparing with the near extremal models. Once we take the near extremal coefficient into our model, our model
yields correct entropy and temperature through all the non-extremal region beyond the near extremal region,
and does not yield any additional discrepancy.

%%%%%%%%%%%%%%%%%%%%%%%%%%%%%%%%%
%
%
%  Conclusion
%
%
%%%%%%%%%%%%%%%%%%%%%%%%%%%%%%%%%

\section{Conclusion and Discussion \label{lastsection}}

\ \ \ \ In this paper, we introduced an ansatz of thermodynamical model
for non-extremal black $p$-brane thermodynamics, which yields the correct entropy 
and temperature and greybody factor when the entropy is maximized.  
We have shown that the geometrical similarity between non-extremal black $p$-brane 
and near extremal black $p$-brane is related to the various properties of the model.
This fact implies that the model is appropriate, and we have actually shown 
the appropriateness of the model for $p=3$ by considering the action of massless scalar field.

Comparing with the models proposed in the past, our model 
can smoothly reach to the near extremal models \cite{B3} \cite{Bp}  
and does not need any additional condition (like equal gas energies on brane and antibrane),
and reproduces correct entropy and temperature through all the
non-extremal region. 

Since the supersymmetry is broken in non-extremal region, the 
definite argument including quantum correction is difficult. Our 
ansatz certainly reproduces the entropy and the temperature of the
non-extremal black $p$-brane, however further study is needed for 
the explanation beyond the geometrical similarity.

\vspace{5ex}

{\bf Acknowledgements}\\
 I would like to thank to Tamiaki Yoneya, Mitsuhiro Kato, Koji Hashimoto, Kenji Hotta, Yuri Aisaka 
for many helpful discussions, and would like to thank to Makoto Natsuume for helpful discussion and advises.

\end{document}